\documentclass[12pt,a4paper,fleqn]{article}

\usepackage[DIV12]{typearea}
\usepackage{amsmath}
\usepackage{amssymb}
\usepackage{amsfonts}
\usepackage{amscd}
\usepackage{graphicx}
\usepackage[footnotesize]{caption2}
\usepackage{cite}
\usepackage[x11names]{xcolor}
\usepackage{bbm}

\newcommand{\eVdist}{\kern-0.06667em}

\newcommand{\ps}{\text{\sc ps}}
\newcommand{\GG}{\text{\sc gg}}

\newcommand{\sm}{\text{\sc sm}}

\def\beqra{\begin{eqnarray}}
\def\eeqra{\end{eqnarray}}

\def\beq{\begin{equation}}
\def\eeq{\end{equation}}

\def\beq{\begin{equation}}
\def\eeq{\end{equation}}
\def\bea{\begin{eqnarray}}
\def\eea{\end{eqnarray}}

\renewcommand{\[}{\left[}

\allowdisplaybreaks[1]

\begin{document}

\begin{titlepage}
\renewcommand{\thefootnote}{\alph{footnote}}

\begin{flushright}
DESY 08-013\\
TUM-HEP 684/08\\
SISSA 16/2008/EP
\end{flushright}

\vspace*{1.0cm}

\renewcommand{\thefootnote}{\fnsymbol{footnote}}

\begin{center}
{\Large\bf Small Extra Dimensions from the Interplay\\ of Gauge and 
Supersymmetry Breaking
} 

\vspace*{1cm}
\renewcommand{\thefootnote}{\alph{footnote}}

\textbf{
Wilfried Buchm\"uller\footnote[1]
{Email: \texttt{wilfried.buchmueller@desy.de}},
Riccardo Catena\footnote[2]{Email: \texttt{catena@sissa.it}}
\\and
Kai Schmidt-Hoberg\footnote[3]{Email: \texttt{kschmidt@ph.tum.de}}
}
\\[5mm]
{\it
$^a$Deutsches Elektronen-Synchrotron DESY, 22603 Hamburg, Germany\\
$^b$ISAS-SISSA, 34013 Trieste, Italy\\
$^c$Physik Department T30, Technische Universit\"at M\"unchen,\\ 85748 Garching, Germany
}

\end{center}

\vspace*{1cm}

\begin{abstract}
\noindent 
Higher-dimensional theories provide a promising framework for unified 
extensions of the supersymmetric standard model. Compactifications to four 
dimensions often lead to $U(1)$ symmetries beyond the standard model
gauge group, whose breaking scale is classically undetermined. Without 
supersymmetry breaking, this is also the case for the size of the compact 
dimensions. Fayet-Iliopoulos terms generically fix the scale $M$ 
of gauge 
symmetry breaking. The interplay with supersymmetry breaking can then 
stabilize the compact dimensions at a size $1/M$, much smaller than the inverse supersymmetry breaking scale $1/\mu$. We illustrate this mechanism 
with an $SO(10)$ model in six dimensions, compactified on an orbifold.
\end{abstract}
\end{titlepage}
\newpage

\section{Introduction}

Higher-dimensional theories provide a promising framework for unified 
extensions of the supersymmetric standard model \cite{wit02}. Interesting
examples have been constructed in five and six dimensions  compactified
on orbifolds \cite{kaw00,af01,hn01,hm01,abc01,hnx01}, which have many 
phenomenologically attractive features. During the past years
it has become clear how to embed these orbifold GUTs into the heterotic string
\cite{krz04,fnx04,bhx04}, separating the GUT scale from the string scale
on anisotropic orbifolds \cite{ht04}. A class of compactifications yielding
supersymmetric standard models in four dimensions (4D) have been successfully 
constructed \cite{bhx05,lnx06,kkk07}.

For a given orbifold compactification of the heterotic string, one can
consider different orbifold GUT limits where one or two of the compact 
dimensions are larger than the other five or four, respectively \cite{bhx04}. 
One then obtains an effective five-dimensional (5D) or six-dimensional (6D)
GUT field theory 
as intermediate step between the full string theory and the supersymmetric
standard model. We shall focus on 6D field theories compactified on  
$T^2/\mathbbm{Z}_2$ with two Wilson lines. These models have four fixed points
where quantum corrections generically induce Fayet-Iliopoulos terms 
\cite{lnz03,bls07}. In the case of the heterotic string the magnitude of these 
local terms is $\mathcal{O}(M_{\rm GUT})$, which suggests that they may lead 
to a stabilization of the compact dimensions at $R\sim 1/M_{\rm GUT}$ 
\cite{bls07}. 

Quantum corrections to the vacuum energy density, the Casimir energy, play
a crucial role in the stabilization of compact dimensions \cite{ac83}. 
Various aspects of the Casimir energy for 6D orbifolds have already been
studied in \cite{pp01,pp03,ghx05}. Stabilization of the volume can be
achieved by means of massive bulk fields, brane localized kinetic terms or
bulk and brane cosmological terms \cite{pp01}. Alternatively, the interplay 
of one- and two-loop contributions to the Casimir energy can lead to a 
stabilization at the length scale of higher-dimensional couplings \cite{gh05}.
In addition, fluxes and gaugino condensates play an important role 
\cite{bht06,hml08}.

In this paper we consider orbifold GUTs, which generically have two
mass scales: $M \sim M_{\rm GUT}$, the expectation value of bulk fields 
induced by local Fayet-Iliopoulos terms,
and $\mu \ll M_{\rm GUT}$, the scale of soft supersymmetry breaking 
mass terms.  As we shall see, the interplay of `classical' and
one-loop contributions to the vacuum energy density can stabilize 
the extra dimensions at small radii, $R \sim 1/M_{\rm GUT} \ll 1/\mu$ with bulk 
energy density ${\cal O}(\mu^2 M^2_{\rm GUT})$. We shall illustrate this 
mechanism with 
an $SO(10)$ model in six dimensions \cite{abc03} which together with 
gaugino mediation \cite{kks99,clx99} is known to lead to a successful
phenomenology \cite{bks05,bcx07}.

The paper is organized as follows. In Section~2 we briefly describe the
relevant features of the 6D orbifold GUT model. The Casimir energies of
scalar fields with different boundary conditions are discussed in Section~3.
These results are used in Section~4 to evaluate the Casimir energy of
the considered model. In Section~5 the stabilization mechanism is described.
Appendices A and B deal with the mode expansion on $T^2/{\mathbbm Z}_2^3$
and the evaluation of Casimir sums, respectively.

\section{The Model}
As an example, we consider a 6D $\mathcal{N}=1$ $SO(10)$ gauge theory 
compactified on an orbifold $T^2/{\mathbbm Z}_2^3$,
corresponding to $T^2/{\mathbbm Z}_2$ with two Wilson lines \cite{abc03}. 
The model has four inequivalent fixed points (`branes') with the unbroken 
gauge groups $SO(10)$, the Pati-Salam group
${G}_{\ps}={ SU(4)}\times { SU(2)} \times {SU(2)}$, the extended Georgi-Glashow
group ${G}_{\GG}={ SU(5)}\times {U(1)}_X$ and flipped $SU(5)$, 
${G}_{\text{fl}}={ SU(5)'}\times {U(1)}'$, respectively. The intersection of 
these GUT groups yields the standard model group with an additional $U(1)$ 
factor, 
${G}_{\sm}'= {SU(3)_C}\times { SU(2)_L} \times { U(1)}_Y\times { U(1)}_{X}$, 
as unbroken gauge symmetry below the compactification scale. At the fixed 
points only 4D $\mathcal{N}=1$ supersymmetry remains unbroken. Gauge and
supersymmetry breaking are realized by assigning different parities to the
different components of the ${\bf 45}$-plet of $SO(10)$, which is a 6D 
$\mathcal{N}=1$ vector multiplet containing 4D $\mathcal{N}=1$ vector ($V$) and
chiral ($\Sigma$) multiplets (cf.~Table~1).

\begin{table}[t]
  \centering
  \renewcommand{\arraystretch}{1.7}
  \begin{tabular}{||l||ccc||c|c||ccc||}
    \hline
    &  
    \multicolumn{5}{c||}{$V$}  &
    \multicolumn{3}{c||}{$\Sigma$} 
    \\
    \cline{2-9}
    ${G_\sm^\prime}$   & ${\mathbbm Z}_2$ & ${\mathbbm Z}_2^\GG$ & ${\mathbbm Z}_2^\ps$ 
    & \multicolumn{2}{c||}{$\mathcal{M}_{m,n}^2$}
    & ${\mathbbm Z}_2$ & ${\mathbbm Z}_2^\GG$ & ${\mathbbm Z}_2^\ps$ \\ 
    \hline\hline
    {$({\bf 8, 1})_{0,0}$}  
    & $+$ & $+$ & $+$ & 
    \multicolumn{2}{c||}{$4\left(\tfrac{m^2}{R_1^2}+\tfrac{n^2}{R_2^2}\right)$} 
    & $-$ & $-$ & $-$   \\
    $({\bf 3, 2})_{-\tfrac{5}{6},0} $
    & $+$ & $+$ & $-$&   
    \multicolumn{2}{c||}{$4\left(\tfrac{m^2}{R_1^2}+\tfrac{(n+1/2)^2}{R_2^2}\right)$} 
    & $-$ & $-$ & $+$  \\
    $({\bf \bar{3}, 2})_{\tfrac{5}{6},0}$ 
    & $+$ & $+$ & $-$ &   
    \multicolumn{2}{c||}{$4\left(\tfrac{m^2}{R_1^2}+\tfrac{(n+1/2)^2}{R_2^2}\right)$} 
    & $-$ & $-$ & $+$ \\
    {$({\bf 1, 3})_{0,0}$} 
    & $+$ & $+$ & $+$ &  
    \multicolumn{2}{c||}{$4\left(\tfrac{m^2}{R_1^2}+\tfrac{n^2}{R_2^2}\right)$}     
    & $-$ & $-$ & $-$ \\
    {$({\bf 1, 1})_{0,0}$} 
    & $+$ & $+$ & $+$ &    
    \multicolumn{2}{c||}{$4\left(\tfrac{m^2}{R_1^2}+\tfrac{n^2}{R_2^2}\right)$}  
    & $-$ & $-$ & $-$ \\
    $({\bf 3, 2})_{\tfrac{1}{6},4}$ 
    & $+$ & $-$ & $-$ &    
    \multicolumn{2}{c||}{$4\left(\tfrac{(m+1/2)^2}{R_1^2}+\tfrac{(n+1/2)^2}{R_2^2}\right)$}  
    & $-$ & $+$ & $+$  \\
    $({\bf \bar{3}; 1})_{-\tfrac{2}{3},4}$ 
    & $+$ & $-$ & $+$ &   
    \multicolumn{2}{c||}{$4\left(\tfrac{(m+1/2)^2}{R_1^2}+\tfrac{n^2}{R_2^2}\right)$}   
    & $-$ & $+$ & $-$ \\
    $({\bf 1, 1})_{1,4}$ 
    & $+$ & $-$ & $+$ &   
    \multicolumn{2}{c||}{$4\left(\tfrac{(m+1/2)^2}{R_1^2}+\tfrac{n^2}{R_2^2}\right)$}  
    & $-$ & $+$ & $-$ \\
    $({\bf \bar{3}, 2})_{-\tfrac{1}{6},-4}$ 
    & $+$ & $-$ & $-$ &   
    \multicolumn{2}{c||}{$4\left(\tfrac{(m+1/2)^2}{R_1^2}+\tfrac{(n+1/2)^2}{R_2^2}\right)$}  
    & $-$ & $+$ & $+$  \\
    $({\bf 3, 1})_{\tfrac{2}{3},-4}$ 
    & $+$ & $-$ & $+$ &   
    \multicolumn{2}{c||}{$4\left(\tfrac{(m+1/2)^2}{R_1^2}+\tfrac{n^2}{R_2^2}\right)$}    
    & $-$ & $+$ & $-$ \\
    $({\bf 1, 1})_{-1,-4}$ 
    & $+$ & $-$ & $+$ &    
    \multicolumn{2}{c||}{$4\left(\tfrac{(m+1/2)^2}{R_1^2}+\tfrac{n^2}{R_2^2}\right)$}   
    & $-$ & $+$ & $-$ \\
    {$({\bf 1, 1})_{0,0}$} 
    & $+$ & $+$ & $+$ &   
    \multicolumn{2}{c||}{$4\left(\tfrac{m^2}{R_1^2}+\tfrac{n^2}{R_2^2}\right)$} 
    & $-$ & $-$ & $-$ \\
    \hline
  \end{tabular}
  \caption{Decomposition of the ${\sf 45}$-plet of $SO(10)$ into multiplets of 
    the extended standard model gauge group 
    $G_\sm^\prime = SU(3)_C\times SU(2)_L\times U(1)_Y \times U(1)_X$
    and corresponding parity assignments. For later convenience we 
    also give the Kaluza-Klein masses $\mathcal{M}_{m,n}^2$.
    \label{tb:adj}}    
\end{table}

The model has three {\bf 16}-plets of matter fields, localized at the 
Pati-Salam, the Georgi-Glashow, and the flipped $SU(5)$ branes. Further,
there are two {\bf 16}-plets, $\phi$ and $\phi^c$ and two {\bf 10}-plets,
$H_5$ and $H_6$ of bulk matter fields. Their mixing with the brane fields 
yields the characteristic flavor structure of the model \cite{abc03,bcx07}.

The Higgs sector consists of two {\bf 16}-plets, $\Phi$ and $\Phi^c$, 
and four {\bf 10}-plets, $H_1,\ldots, H_4$, of bulk hypermultiplets.
Each hypermultiplet contains two 4D $\mathcal{N}=1$ chiral multiplets, the
first of which we denote by the same symbol as the hypermultiplet.   
The Higgs multiplets have even $R$-charge and the matter fields have odd 
$R$-charge.

The hyperpermultiplets $H_1$ and $H_2$ contain
the two Higgs doublets of the supersymmetric standard model as zero modes,
whereas the zero modes of $H_3$ and $H_4$ are color triplets (cf.~Table~2).
The zero modes of the {\bf 16}-plets are singlets and color triplets,
\begin{align}
&\Phi:\;\; N^c,\ D^c\;;\qquad \Phi^c:\;\; N,\ D\;.
\end{align}
The color triplets $D^c$ and $D$, together with the zero modes of $H_3$ and
$H_4$, aquire masses through brane couplings.

Equal vacuum expectation values of $\Phi$ and $\Phi^c$ form a flat direction 
of the classical potential,
\begin{align}\label{vev}
&\langle \Phi \rangle = \langle N^c \rangle 
= \langle N \rangle = \langle \Phi^c \rangle \;.
\end{align} 
Non-zero expectation values can be enforced by a brane superpotential term or 
by a Fayet-Iliopoulos term localized at the GG-brane where the $U(1)$ factor 
commutes with the standard model gauge group.

The expectation values (\ref{vev})
 break $SO(10) \rightarrow SU(5)$, and therefore also
the additional $U(1)_X$ symmetry, as is clear from the decomposition 
\begin{eqnarray}
{\bf 16} &\rightarrow & 
{\bf 10}_{1} \oplus \bar{{\bf 5}}_{-3} \oplus {\bf 1}_{5}\;, \label{16-1} \\
\overline{{\bf 16}} &\rightarrow & 
\overline{{\bf 10}}_{-1} \oplus {\bf 5}_{3} \oplus {\bf 1}_{-5} \label{16-2}\;, \end{eqnarray}
where ${\bf 1}_{5}$ and  ${\bf 1}_{-5}$ correspond to $N^c$ and $N$, 
respectively. The decomposition of the ${\bf 45}$  vector multiplet reads
\begin{equation}
{\bf 45} \rightarrow {\bf 24}_{0} \oplus {\bf 10}_{-4} 
\oplus \overline{{\bf 10}}_{4}\oplus {\bf 1}_{0}\;. \label{45}
\end{equation}
The expectation values (\ref{vev}) generate for the ${\bf 10}$- and 
$\overline{\bf 10}$-plets and the singlet in Eqs.~(\ref{16-1})-(\ref{45}) the 
bulk mass
\begin{equation}\label{higgsmass}
M^2 \simeq  g_6^2 \langle \Phi^c \rangle^2\;,
\end{equation}
where $g_6$ is the 6D gauge coupling. 
Hence, the fields 
$({\bf 3,2})_{\frac{1}{6},4}$, $({\bf \bar{3},1})_{-\frac{2}{3},4}$, 
$({\bf 1,1})_{1,4}$, $({\bf 1,1})_{0,0}$  
and their complex conjugates contained in the vector 
multiplet as well as the corresponding fields in $\Phi$ and $\Phi^c$
obtain bulk masses from the Higgs mechanism in 
addition to their Kaluza-Klein masses.
Since the spontaneous breaking of $SO(10)$ preserves 6D $\mathcal{N}=1$
supersymmetry, one obtains an entire massive hypermultiplet for each set of 
quantum numbers. 

\begin{table}[t]
  \centering
  \renewcommand{\arraystretch}{1.3}
  \begin{tabular}{|c||cc|cc|cc|cc|}
     \hline
    $SO(10)$ & \multicolumn{8}{c|}{\bf 10} \\
    \hline
    {\sf SM}$^{\prime}$ & \multicolumn{2}{c|}{$(\mathbf{1,2})_{-\tfrac{1}{2},-2}$} &
    \multicolumn{2}{c|}{$(\mathbf{1,2})_{\tfrac{1}{2},2}$} &
    \multicolumn{2}{c|}{$(\mathbf{\bar{3},1})_{\tfrac{1}{3},-2}$} 
    & \multicolumn{2}{c|}{$(\mathbf{3,1})_{-\tfrac{1}{3},2}$} \\ 
    \hline
    & \multicolumn{2}{c|}{$H^c$} & \multicolumn{2}{c|}{$H$} &
    \multicolumn{2}{c|}{$G^c$} & \multicolumn{2}{c|}{$G$} \\
    & ${\mathbbm Z}_2^\ps$ & ${\mathbbm Z}_2^\GG$ & ${\mathbbm
      Z}_2^\ps$ & ${\mathbbm Z}_2^\GG$ & ${\mathbbm Z}_2^\ps$ &
    ${\mathbbm Z}_2^\GG$ & ${\mathbbm Z}_2^\ps$ & ${\mathbbm Z}_2^\GG$
    \\
    \hline
    \hline
    $H_1$ & $+$ & $+$ & $+$ & $-$ & $-$ & $+$ & $-$ & $-$ \\
    \hline
    $H_2$ & $+$ & $-$ & $+$ & $+$ & $-$ & $-$ & $-$ & $+$ \\
    \hline
    $H_3$ & $-$ & $+$ & $-$ & $-$ & $+$ & $+$ & $+$ & $-$ \\
    \hline
    $H_4$ & $-$ & $-$ & $-$ & $+$ & $+$ & $-$ & $+$ & $+$ \\
    \hline
    $H_5$ & $-$ & $+$ & $-$ & $-$ & $+$ & $+$ & $+$ & $-$ \\
    \hline
    $H_6$ & $-$ & $-$ & $-$ & $+$ & $+$ & $-$ & $+$ & $+$ \\
    \hline
    \hline
    $SO(10)$ & \multicolumn{8}{c|}{\bf 16} \\
    \hline
    {\sf SM}$^{\prime}$ & \multicolumn{2}{c|}{$({\bf 3,2})_{\tfrac{1}{6},-1}$} &
    \multicolumn{2}{c|}{$({\bf 1,2})_{-\tfrac{1}{2},3}$} &
    \multicolumn{2}{c|}{$(\mathbf{\bar{3},1})_{-\tfrac{2}{3},-1}$}
    & \multicolumn{2}{l|}{$(\mathbf{\bar{3},1})_{\tfrac{1}{3},3}$} \\ 
    & \multicolumn{2}{|c|}{} &
    \multicolumn{2}{c|}{} & \multicolumn{2}{l|}{$(\mathbf{1,1})_{1,-1}$} 
    & \multicolumn{2}{l|}{$(\mathbf{1,1})_{0,-5}$}
   \\ 
    \hline
    & \multicolumn{2}{c|}{$Q$} & \multicolumn{2}{c|}{$L$} &
    \multicolumn{2}{c|}{$U^c,E^c$} &
    \multicolumn{2}{c|}{$D^c,N^c$} 
    \\ 
    & ${\mathbbm Z}_2^\ps$ & ${\mathbbm Z}_2^\GG$ & ${\mathbbm
      Z}_2^\ps$ & ${\mathbbm Z}_2^\GG$ & ${\mathbbm Z}_2^\ps$ &
    ${\mathbbm Z}_2^\GG$ & ${\mathbbm Z}_2^\ps$ & ${\mathbbm Z}_2^\GG$
    \\
    \hline
    $\Phi$ & $-$ & $-$ & $-$ & $+$ & $+$ & $-$ & $+$ & $+$ \\
    \hline 
    $\phi$ & $+$ & $-$ & $+$ & $+$ & $-$ & $-$ & $-$ & $+$ \\
    \hline  
  \end{tabular}
  \caption{Decomposition and parity assignments for the bulk ${\bf 16}$- 
           and ${\bf 10}$-plets of $SO(10)$. The 
           ${\mathbf {\overline{16}}}$-plets $\Phi^c,\phi^c$ have the same 
           parities as $\Phi$ and $\phi$ and conjugate quantum 
           numbers with respect to
           the extended standard model gauge group. Only fields with
           all parities positive remain in the low energy theory.
    \label{dec}}
\end{table}

Supersymmetry breaking is naturally incorporated via gaugino mediation 
\cite{bks05}. The non-vanishing $F$-term of a brane field $S$ generates
mass terms for vector- and hypermultiplets. In the considered model, $S$
is localized at the $SO(10)$ preserving brane, which yields the same mass
for all members of an $SO(10)$ multiplet. For the ${\bf 45}$ vector multiplet
and the ${\bf 10}$ and {\bf 16} hypermultiplets of the Higgs sector one has
\newpage
\begin{eqnarray}\label{mass}
\Delta S &=& \int \text{d}^4x \text{d}^2y\ \delta^2(y) \left\{\int \text{d}^2 \theta  
 \frac{1}{2\Lambda^3} S {\rm Tr}[W^\alpha W_\alpha] + {\rm h.c.} 
 \right. \nonumber\\ 
&& \left. + \int \text{d}^4\theta \left(
\frac{\lambda}{\Lambda^4} S^\dagger S 
  \left(H_1^\dagger H_1 + H_2^\dagger H_2\right)  
 + \frac{\lambda'}{\Lambda^4} S^\dagger S 
  \left(H_3^\dagger H_3 + H_4^\dagger H_4\right)\right.\right.\nonumber\\ 
&& \hspace{16mm} \left.\left. + \frac{\lambda''}{\Lambda^4} S^\dagger S 
  \left(\Phi^\dagger \Phi + \Phi^{c\dagger} \Phi^c\right) \right)\right\}\;.
\end{eqnarray}
Here $W^\alpha(V)$, $H_1,\ldots, H_4$ and $\Phi, \Phi^c$ are the 4D 
$\mathcal{N}=1$ multiplets
contained in the 6D $\mathcal{N}=1$ multiplets, which have positive parity
at $y=0$; $\Lambda$ is the UV cutoff of the model, which is much larger 
than the inverse size of the compact dimensions, $\Lambda \gg 1/\sqrt{V}$.
For the zero modes, the corresponding gaugino and scalar masses are given by
\begin{equation}\label{masses}
m_g = \frac{\mu}{\Lambda^2 V}\;, \quad 
m^2_{H_{1,2}} = - \frac{\lambda \mu^2}{\Lambda^2 V}\;, \quad
m^2_{H_{3,4}} = - \frac{\lambda' \mu^2}{\Lambda^2 V}\;, \quad
m^2_{\Phi} = - \frac{\lambda'' \mu^2}{\Lambda^2 V}\;,  
\end{equation} 
where $V = (2\pi)^2 R_1R_2$ is the volume of the compact dimensions, and
$\mu=F_S/\Lambda$. Note that the gaugino mass is stronger volume 
suppressed than the scalar masses.

\section{The Casimir Energy}

The zero-point energies of bulk fields depend on size and shape of the
compact dimensions. Their sum, the Casimir energy, is a quantum contribution
to the total energy density whose minimum determines the size of the
compact dimensions in the lowest energy state, the vacuum. As long as 
supersymmetry is unbroken, the Casimir energy vanishes since bosonic and
fermionic contributions compensate each other.
In the following we shall evaluate the Casimir energy for the different
boundary conditions which occur in $T^2/\mathbbm{Z}_2^3$ orbifold 
compactifications.

\subsection{Bulk, Brane and Kaluza-Klein Masses}

Consider a real scalar field in 6D with bulk mass $M$ and brane mass $m$.
As discussed in the previous section, in gaugino mediation $m$ is due to 
supersymmetry breaking on a brane whereas $M$ is generated by the Higgs 
mechanism in 6D. From the action
\begin{align}
S &= \frac{1}{2} \int \text{d}^4x \text{d}^2y \phi(x,y)\left(-\partial_x^2 - \partial_y^2
+ M^2 + \frac{\mu^2}{\Lambda^2} \delta^2(y)\right)\phi(x,y)
\end{align}
and the mode decomposition
\begin{align}
\phi(x,y) = \sum_i \phi_i(x) \xi_i(y)\;, \quad
\int d^2y \xi_i(y)\xi_j(y) = \delta_{ij}\;,
\end{align}
one obtains
\begin{align}
S &= \frac{1}{2} \int \text{d}^4x \left[\sum_{i}\phi_i(x)
\left(-\partial_x^2 + M_i^2 + M^2\right)\phi_i(x) + 
\frac{\mu^2}{\Lambda^2} \sum_{ij} \phi_i(x) C_{ij} \phi_j(x)\right]\;,
\end{align}
where $M_i$ are the Kaluza-Klein masses and
\begin{equation}
C_{ij} = \xi_i(0)\xi_j(0)\;.
\end{equation}
On the orbifold $T^2/\mathbbm{Z}_2^3$, one has for all modes (cf.~Appendix~A),
\begin{equation}
\xi_i(0) = \sqrt{\frac{2}{V}} = \frac{1}{\sqrt{2\pi^2R_1R_2}}\;,
\end{equation}
except for the zero mode, where $\xi_0(y) = 1/\sqrt{V}$.
 
The one-loop contribution to the vacuum energy density depends on the 
Kaluza-Klein mass matrix $M_{KK}$, the universal mass $M$ and the brane
mass matrix $C$,
\begin{align}
V^{(1)} &= \frac{1}{2}\ln\det\left(-\partial_x^2 + M^2_{\rm KK} + M^2 + 
\frac{\mu^2}{\Lambda^2} C\right)\;.
\end{align}
For small supersymmetry breaking, $\mu^2 \ll M_i^2 + M^2$, the effective 
potential can be expanded in powers of the small off-diagonal terms of the
mass matrix,
\begin{align}
V^{(1)} &= \frac{1}{2}\sum_i\ln\left(-\partial_x^2 + M^2_i + M^2 + 
\frac{\mu^2}{\Lambda^2} C_{ii}\right) \nonumber \\
&+\frac{1}{2} \left(\frac{\mu^2}{\Lambda^2}\right)^2  
\sum_{i\neq j} \frac{1}{(-\partial_x^2 + M^2_i + M^2)} 
C_{ij} \frac{1}{(-\partial_x^2 + M^2_j + M^2)} C_{ji}
+ {\cal O}(\mu^6) \;.
\end{align}
In the following we shall only keep the diagonal terms of $C$, which contribute
to $V^{(1)}$ at leading order in $\mu^2$.

The Casimir energy of gauge fields and gauginos can be directly obtained from 
the Casimir energy of a real scalar field. After appropriate gauge fixing this
essentially amounts to counting the physical degrees of freedom 
(cf.~\cite{pp01}). Thus, it is enough to perform the vacuum energy calculation
for a real scalar field.

\subsection{Casimir Energy of a Scalar Field}

The geometry of the orbifold $T^2/\mathbbm{Z}_2$ contains as free
parameters the radii $R_1$ and $R_2$ of the torus. The Casimir energy
of a scalar field on the orbifold is then given by the quantum corrections
to the corresponding effective potential. At one-loop order, this is obtained 
by summing over the continuous and discrete spectrum corresponding to the four 
flat and two compact dimensions,  
\begin{align}\label{effpot}
V_M
&= \frac{1}{2} \left[
\sum \right]_{m,n}
\int \frac{\text{d}^4k_E}{(2\pi)^4} 
\log\left(k_E^2 +\mathcal{M}^2_{m,n}+M^2\right) 
\;,
\end{align}
with $\left[\sum \right]_{m,n}$ shorthand for the double sum
and $\mathcal{M}^2_{m,n}$ denoting the Kaluza-Klein masses; the mass
$M$ now stands for bulk and brane mass terms.

The Kaluza-Klein masses $\mathcal{M}^2_{m,n}$ depend on the possible boundary 
conditions on $T^2/\mathbbm{Z}_2$ and can be read off from the mode expansion
listed in Table~\ref{tb:adj}. Generically they can be written as 
\begin{align}
\mathcal{M}^2_{m,n} &= 4 \left[\frac{(m+\alpha)^2}{R_1^2}
+\frac{(n+\beta)^2}{R_2^2}\right] \nonumber \\
&= \frac{4}{R_2^2} \left[e^2(m+\alpha)^2 +(n+\beta)^2\right] \;,
\end{align}
where $(\alpha,\beta) = (0,0), (0,1/2), (1/2,0), (1/2,1/2)$ and 
$e^2 = R_2^2/R_1^2$. For simplicity, we restrict our discussion
to `rectangular tori'. The general case will be discussed elsewhere 
\cite{bcs08}.
Clearly, the contributions for the different boundary conditions satisfy 
the relations,
\begin{align}\label{symmetry}
V_M^{0,0}(R_1,R_2) &= V_M^{0,0}(R_2,R_1)\;,\quad
V_M^{1/2,1/2}(R_1,R_2) = V_M^{1/2,1/2}(R_2,R_1)\;,\nonumber\\
V_M^{0,1/2}(R_1,R_2) &= V_M^{1/2,0}(R_2,R_1)\;.
\end{align}

The expression (\ref{effpot}) for the Casimir energy is divergent. Following
\cite{eli94,pp01}, we extract a finite piece using zeta function 
regularization, 
\begin{align}
V
&= - \frac{\text{d}\zeta(s)}{\text{d}s}\bigg|_{s=0} \;,
\end{align}
where 
\begin{align}\label{zeta}
\zeta(s) &= 
\frac{1}{2} \left[
\sum \right]_{m,n} \mu_r^{2s}
\int \frac{\text{d}^4k_E}{(2\pi)^4} 
\left(k_E^2 +  \frac{4}{R_2^2} \left[e^2(m+\alpha)^2
+(n+\beta)^2\right]  +   M^2\right)^{-s}\;.
\end{align}
Note that, as in dimensional regularization, a mass scale $\mu_r$ is 
introduced. The momentum integration can now be carried out and one obtains 
\begin{align}\label{zeta}
\zeta(s)
&= \frac{1}{2} \frac{1}{(2\pi)^4} 
\pi^2 \frac{\Gamma(s-2)}{\Gamma(s)} \left[\sum \right]_{m,n}\mu_r^{2s}
\left( \frac{4}{R_2^2} \left[e^2(m+\alpha)^2
+(n+\beta)^2\right]  +   M^2\right)^{2-s} \nonumber \\
&= \frac{4^{2-s}}{32\pi^2R_2^{4-2s}}
\frac{\mu_r^{2s}}{(s-2)(s-1)} \left[\sum \right]_{m,n}
\left( \left[e^2(m+\alpha)^2
+(n+\beta)^2\right]  +  \frac{R_2^2}{4}  M^2\right)^{2-s}  .
\end{align}

The boundary conditions of fields on the orbifold $T^2/\mathbbm{Z}_2^3$
are characterized by three parities. For positive (negative) parity the field
is nonzero (zero) at the corresponding fixed point. For the Casimir energy
only those chiral and vector multiplets are relevant which are nonzero at 
the fixed point where supersymmetry is broken. Hence one parity, chosen
to be the first one, has to be positive. Inspection of the mode expansion
in Appendix~A shows that for the fields  $\Phi_{+-\pm}$, corresponding to
$(\alpha,\beta)=(1/2,0),(1/2,1/2)$, one has to perform the sum
\begin{align}\label{sum1}
 \left[\sum \right]_{m,n} = 
\sum_{m=0}^{\infty}\sum_{n=-\infty}^{\infty} \;,
\end{align} 
whereas the boundary conditions $\Phi_{++\pm}$, with
$(\alpha,\beta)=(0,0),(0,1/2)$, requires the sum 
\begin{align}\label{sum2}
 \left[\sum \right]_{m,n} = 
 \left[ \delta_{0,m} \sum_{n=0}^{\infty}+
\sum_{m=1}^{\infty}\sum_{n=-\infty}^{\infty} \right] \;.
\end{align}

The two summations (\ref{sum1}) and (\ref{sum2}) are carried out in Appendix~B.
The result can be expressed in the following form, which is suitable for 
numerical analysis,
\begin{align}\label{double}
V_M^{\alpha\beta}(R_1,R_2)
=&\; \frac{M^6 R_1 R_2}{768\pi}\left(\frac{11}{12} - 
              \log\left(\frac{M}{\mu_r}\right)\right)\nonumber\\
  & - \delta_{\alpha0}\delta_{\beta0}\frac{M^4}{64\pi^2}\left(\frac{3}{4} - 
                \log\left(\frac{M}{\mu_r}\right)\right)\nonumber \\
& - \frac{1}{8\pi^4} \frac{M^3 R_2}{R_1^2}
\sum_{p=1}^{\infty}\frac{\cos(2\pi p \alpha)}{p^3}
K_{3}(\pi p M R_1) \nonumber \\
& -  \frac{2}{\pi^4} \frac{1}{R_2^4}
\sum_{p=1}^{\infty} \frac{\cos(2 \pi p \beta)}{p^{5/2}} 
 \sum_{m=0}^{\infty}\frac{1}{2^{\delta_{\alpha0}\delta_{m0}}}
\left(\tfrac{R_2}{R_1}\sqrt{(m+\alpha)^2 +
         \tfrac{M^2R_1^2}{4}}\right)^{5/2} \nonumber \\
&\hspace{2.5cm}K_{5/2} \left(2\pi \,p \,\tfrac{R_2}{R_1}
\sqrt{(m+\alpha)^2 + \tfrac{M^2R_1^2}{4}}\right) \;.
\end{align}
We have checked numerically that this expression satisfies the symmetry
relations (\ref{symmetry}). As a good approximation, where the symmetries 
are manifest, one can derive \cite{bcs08}
\begin{align}\label{nice}
V_M^{\alpha\beta}(R_1,R_2)
=&\; \frac{M^6 R_1 R_2}{768\pi}\left(\frac{11}{12} - 
              \log\left(\frac{M}{\mu_r}\right)\right)\nonumber\\
  & - \delta_{\alpha0}\delta_{\beta0}\frac{M^4}{64\pi^2}\left(\frac{3}{4} - 
                \log\left(\frac{M}{\mu_r}\right)\right)\nonumber \\
& - \frac{1}{8\pi^4} \frac{M^3 R_2}{R_1^2}
\sum_{p=1}^{\infty}\frac{\cos(2\pi p \alpha)}{p^3}
K_{3}(\pi p M R_1) 
 \nonumber \\
& - \frac{1}{8\pi^4} \frac{M^3 R_1}{R_2^2}
\sum_{p=1}^{\infty}\frac{\cos(2\pi p \beta)}{p^3}K_{3}(\pi p M R_2) \;.
\end{align}
The term $\propto \delta_{\alpha 0} \delta_{\beta 0}$, which is independent 
of $R_1$ and $R_2$, is precisely the contribution of the `zero' mode
in (\ref{zeta}), with $\alpha=\beta=m=n=0$.  

\begin{figure}
  \centering
  \begin{minipage}[b]{7.4 cm}
    \includegraphics[width=7.4cm,height=6cm]{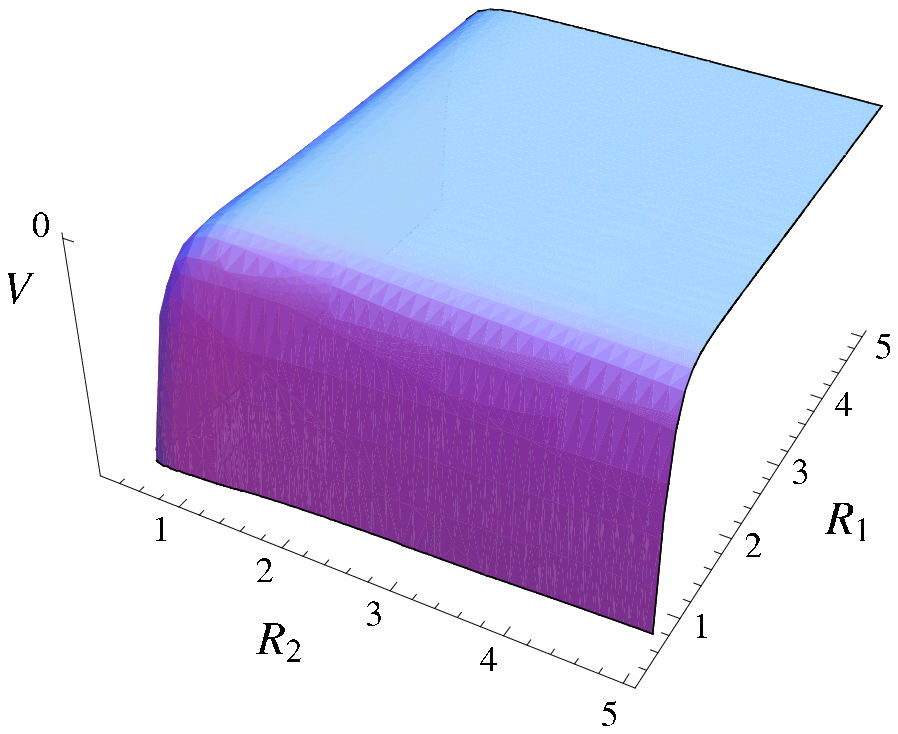}  
  \end{minipage}\hspace*{1cm}
  \begin{minipage}[b]{7.4 cm}
    \includegraphics[width=7.4cm,height=6cm]{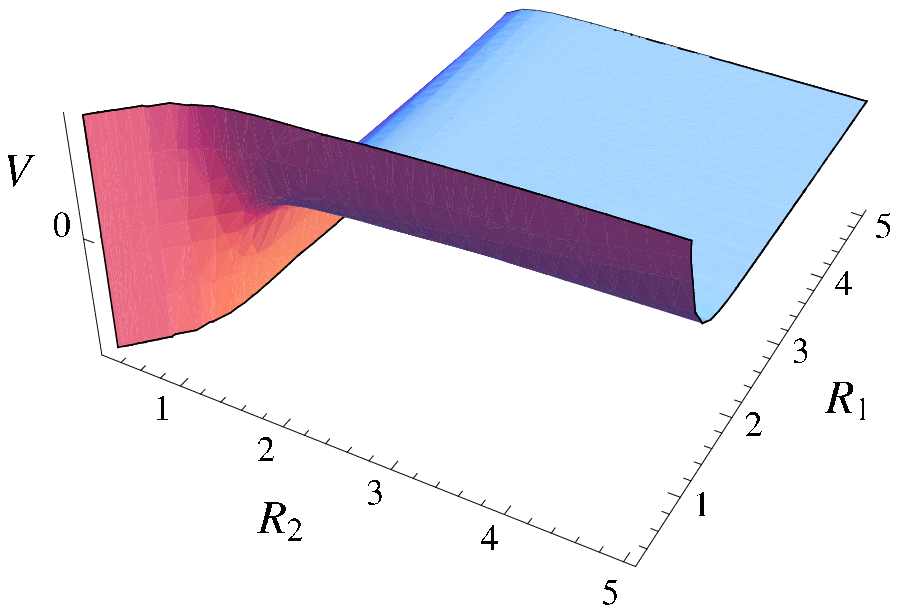}  
  \end{minipage}\vspace*{1cm}
  \begin{minipage}[b]{7.4 cm}
    \includegraphics[width=7.4cm,height=6cm]{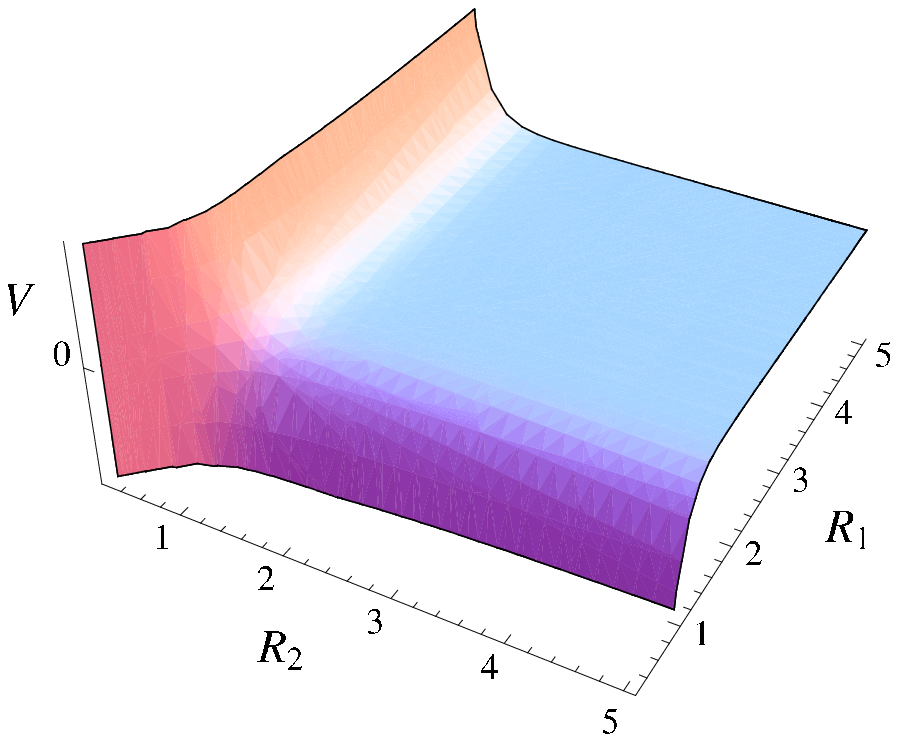}  
  \end{minipage}\hspace*{1cm}
  \begin{minipage}[b]{7.4 cm}
    \includegraphics[width=7.4cm,height=6cm]{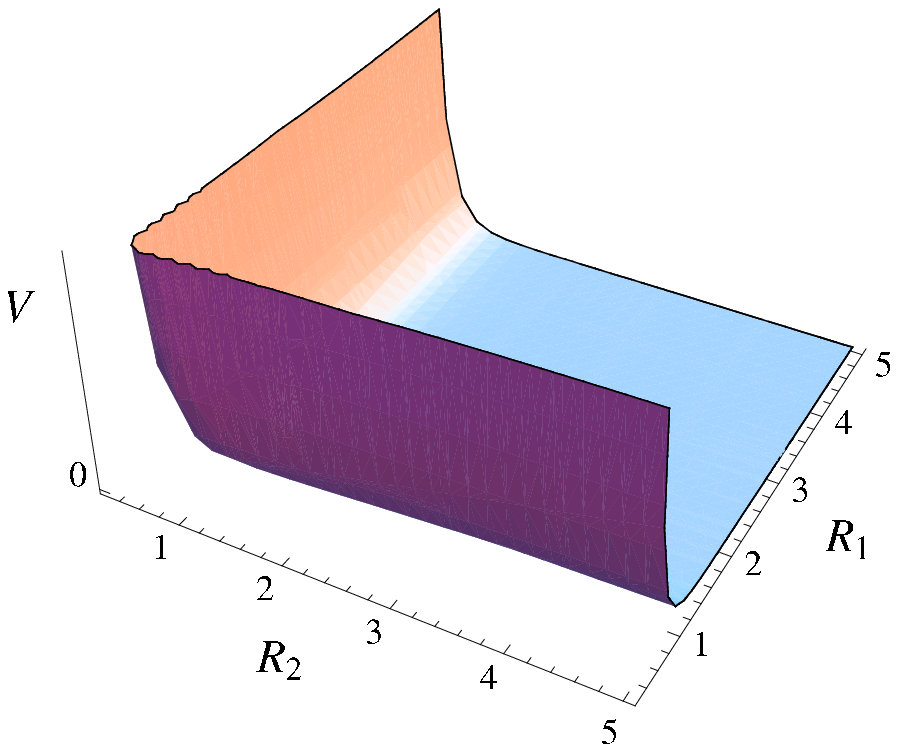}  
  \end{minipage}
  \caption{The four different contributions to the Casimir energy in units of
           the supersymmetric mass $M$.
           From top left to bottom right we have $V_M^{0,0}$, $V_M^{1/2,0}$,
           $V_M^{0,1/2}$ and $V_M^{1/2,1/2}$ as defined in the text.}
  \label{fig:V}
\end{figure}

The dependence of the first two terms in (\ref{double})
on the regularization scale $\mu_r$ is a remnant of the subtraction
of divergent bulk and brane cosmological terms, as in dimensional 
regularization \cite{ghx05}. The corresponding contributions to the anomalous 
dimensions of the 6D and 4D cosmological terms read
\begin{equation}
\gamma_6 = \mu_r \frac{\partial}{\partial\mu_r} \Lambda_6 
= - \frac{M^6 R_1R_2}{768\pi}\;,\quad
\gamma_4 = \mu_r \frac{\partial}{\partial\mu_r} \Lambda_4 
= \frac{M^4}{64\pi}\;.
\end{equation}
The presence of these terms demonstrates that the renormalization of the
divergent energy density (\ref{double}) requires counter terms for the bulk 
and brane cosmological terms.

In general, the Casimir energy is a sum of the four possible terms,
\begin{equation}\label{nonsusy}
V_M =
A  V_M^{0,0} + B V_M^{0,1/2} + C V_M^{1/2,0} + D V_M^{1/2,1/2}\;,
\end{equation}
where the coefficients $A$,...,$D$ depend on the field content of the model
and we have assumed equal masses for simplicity. The four functions 
$V_M^{0,0}$,...,$V_M^{1/2,1/2}$ are shown in Figure~1. For small $R_{1,2}$,
$V_M^{0,0}$ is attractive and $V_M^{1/2,1/2}$ is repulsive, whereas the
other two have mixed behavior.

In supersymmetric theories there is a cancellation between bosonic and 
fermionic contributions, and the expression (\ref{nonsusy}) for the
Casimir energy is replaced by
\begin{eqnarray}\label{susy}
V&=&
  a \left( V_{M'}^{0,0}- V_M^{0,0}\right)
+ b \left( V_{M'}^{0,1/2}- V_M^{0,1/2}\right) 
\nonumber \\
&&+c \left( V_{M'}^{1/2,0} -V_M^{1/2,0}\right)
+ d \left( V_{M'}^{1/2,1/2}-V_M^{1/2,1/2}\right)\;,
\end{eqnarray}
where $M'=\sqrt{M^2+m^2}$, with supersymmetric mass $M$ and  
supersymmetry breaking mass $m$; the coefficients $a$,...,$d$ again depend 
on the field content of the model.
Compared to the non-supersymmetric case (\ref{nonsusy}), the behavior at 
small $R_{1,2}$ is inverted. For bulk vector- and hypermultiplets only the
4D $\mathcal{N}=1$ vector and chiral multiplets are relevant, which couple
to the brane where supersymmetry is broken.

The qualitative behavior of Figure~1 is easily understood by evaluating
explicitly the Casimir energy (\ref{nice}) at small radii 
$R_1, R_2 \ll 1/M$. 
Expanding the Bessel function $K_3$ for small arguments and performing the 
summations over $p$, one obtains
\begin{eqnarray}\label{expansion}
 V_M^{0,0}(R_1,R_2) 
&=& -\frac{1}{945\pi}\frac{R_2}{R_1^5}\left(1 - \frac{21}{16}M^2 R_1^2
+ \ldots \right)\ + \ R_1 \leftrightarrow R_2 \;, \\
V_M^{0,1/2}(R_1,R_2) 
&=& -\frac{1}{945\pi}\frac{R_2}{R_1^5}\left(1 - \frac{21}{16}M^2 R_1^2
+ \ldots \right) \nonumber\\
&&+ \frac{31}{30240\pi}\frac{R_1}{R_2 ^5}\left(1 - 
\frac{147}{124}M^2 R_2^2 + \ldots \right)\;,\\
V_M^{1/2,0}(R_1,R_2) 
&=& \frac{31}{30240\pi}\frac{R_2}{R_1^5}\left(1 - \frac{147}{124}M^2 R_1^2
+ \ldots \right) \nonumber\\
&&- \frac{1}{945\pi}\frac{R_1}{R_2 ^5}\left(1 - 
\frac{21}{16}M^2 R_2^2 + \ldots \right)\;,\\
V_M^{1/2,1/2}(R_1,R_2) 
&=& \frac{31}{30240\pi}\frac{R_2}{R_1^5}\left(1 - \frac{147}{124}M^2 R_1^2
+ \ldots \right)\ +\ R_1 \leftrightarrow R_2 \;.  
\end{eqnarray}
From these equations one immediately reads off the behavior of 
$V_M^{\alpha,\beta}$ at small radii. For $R_{1,2} \rightarrow 0$, with
$R_1/R_2$ fixed, one obtains the behavior of the Casimir energy
for 5D orbifolds. For supersymmetric models, the mass independent terms
cancel, and with $M'^2-M^2=m^2$ the second terms in the expansion yield
the inverted behavior at small $R_{1,2}$.

\section{Casimir Energy of the Orbifold Model}

Given the results of the previous section we can now easily evaluate the
Casimir energy of the orbifold GUT model described in Section~2. At the
branes, only 4D $\mathcal{N}=1$ supersymmetry is preserved. A multiplet 
contributes to the Casimir energy if its bosonic and fermionic degrees 
of freedom have different masses. This only happens if its first 
${\mathbbm Z}_2$ parity is positive so that it can couple to the singlet $S$ 
at the $SO(10)$ brane, whose non-vanishing $F$-term breaks 4D $\mathcal{N}=1$ 
supersymmetry spontaneously. Hence, from the 6D $\mathcal{N}=1$ vector 
multiplet only $V$ contributes (cf.~Table~1). Also for the hypermultiplets
only one 4D $\mathcal{N}=1$ chiral multiplet is relevant. The corresponding 
chiral multiplets with positive ${\mathbbm Z}_2$ parity are listed in Table~2.

\subsection{Contribution from the Vector Multiplet}

The expectation values (\ref{vev}) break $SO(10)$ spontaneously to $SU(5)$.
This generates the mass $M$ for the 21 vector multiplets of the coset 
$SO(10)/SU(5)$\footnote{We shall ignore the $\mathcal{O}(1)$ factors for the masses of 
different $SU(5)$ representations as they will not be important in the
following discussion.}. Since the Higgs mechanism preserves 6D $\mathcal{N}=2$
supersymmetry, also 21 hypermultiplets become massive.
In addition all gauginos acquire a supersymmetry breaking mass $m_g$.

From Tables \ref{tb:adj} and \ref{dec} and from the mode decomposition
we can now read off the total Casimir energy of the massive vector multiplet
on $T^2/{\mathbbm Z}_2^3$, 
\begin{eqnarray}\label{vector}
V_g &=& 
24 \left(V^{0,0}- V_{m_g}^{0,0}\right) 
+ 24 \left(V^{0,1/2}- V_{m_g}^{0,1/2}\right) 
+ 2 \left(V_M^{0,0}- V_{M'}^{0,0}\right)
\nonumber \\
&& + 16 \left(V_M^{1/2,0}- V_{M'}^{1/2,0}\right)
+ 24 \left(V_M^{1/2,1/2}- V_{M'}^{1/2,1/2}\right)\;,
\end{eqnarray}
where $M'=\sqrt{M^2+m_g^2}$. 
Using the expansion (\ref{expansion}) and $m_g = \mu/(\Lambda^2 V)$ one finds
at small radii,
\begin{equation}
V_g = - \frac{1}{48\pi}\frac{\mu^2}{\Lambda^4 V^2}\left(\frac{R_2}{R_1^3}
+ \ldots \right)\;,
\end{equation}
where the dots denote terms of relative order $\mathcal{O}(M_i R_{1,2})$,
with $M_i = m_g, M, M'$, which have been neglected.

\begin{figure}
  \centering
  \begin{minipage}[b]{7.8 cm}
    \includegraphics[width=7.8cm,height=6cm]{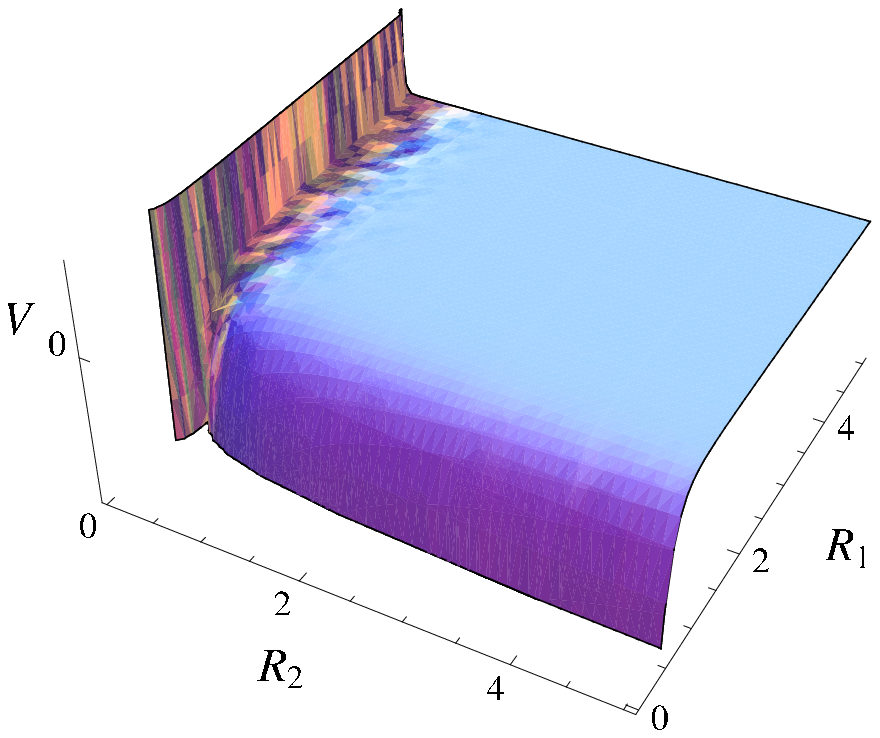}  
  \end{minipage}\hspace*{1cm}
  \begin{minipage}[b]{7.8 cm}
    \includegraphics[width=7.8cm,height=6cm]{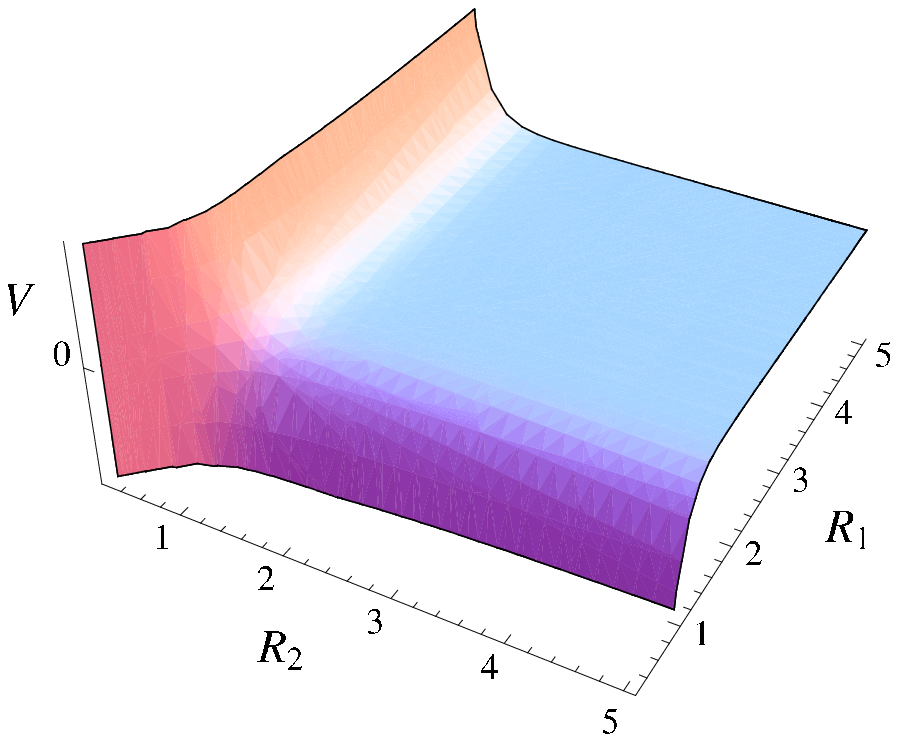}  
  \end{minipage}\vspace*{1cm}
  \begin{minipage}[b]{7.8 cm}
    \includegraphics[width=7.8cm,height=6cm]{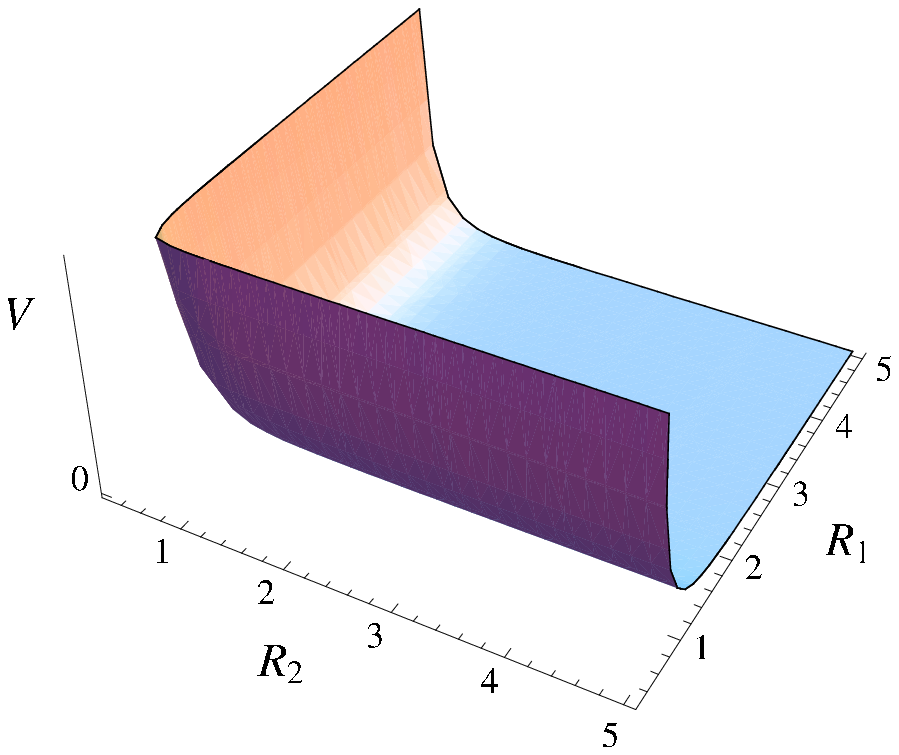}  
  \end{minipage}\hspace*{1cm}
  \begin{minipage}[b]{7.8 cm}
    \includegraphics[width=7.8cm,height=6cm]{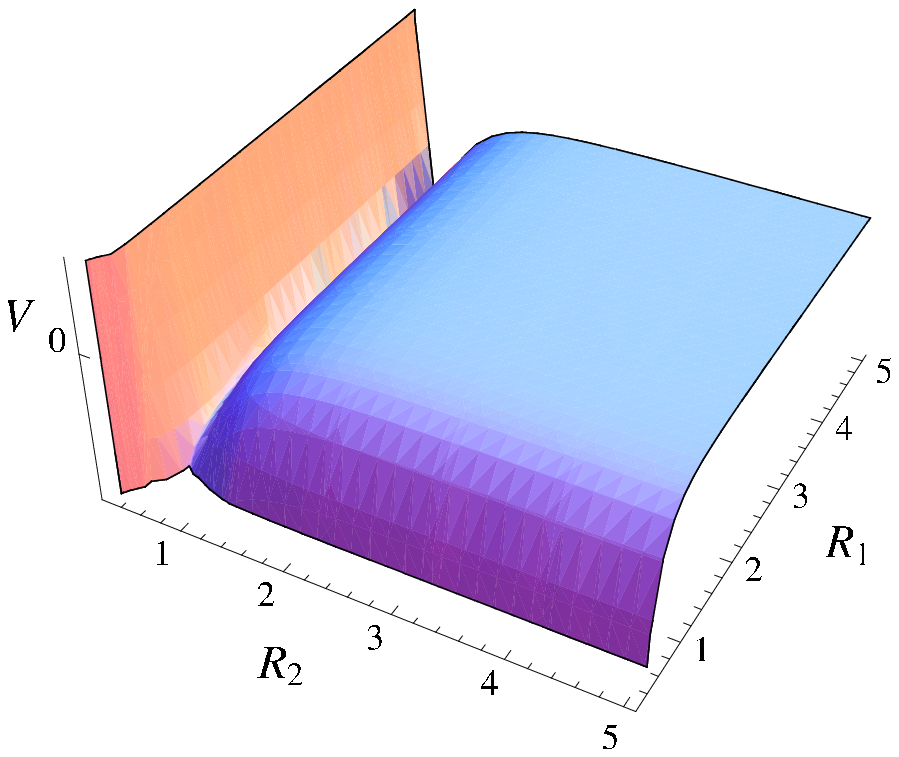}  
  \end{minipage}
  \caption{The different contributions to the Casimir energy from the bulk
  vector multiplet and the hypermultiplets of the Higgs sector (see text).
  From top left to bottom right we have the contributions from the vector
  multiplet, the $\bf 10$-plets $H_{1,2}$, the $\bf 10$-plets $H_{3,4}$, 
  and the $\bf 16$-plets $\Phi,\Phi^c$.}
  \label{fig:Vs}
\end{figure}

\subsection{Contributions from Hypermultiplets}

The contribution of hypermultiplets to the Casimir energy again depends on the
symmetry breaking, i.e., the choice of parities. Consider the {\bf 10}-plets
$H_{1,2}$ which contain the Higgs doublets as zero mode. From Table~2
one reads off, 
\begin{eqnarray}
V_H &=& 
  8 \left(V_{m_H}^{0,0}- V^{0,0}\right)
+ 8 \left(V_{m_H}^{0,1/2}- V^{0,1/2}\right) \nonumber \\
&&+ 12 \left(V_{m_H}^{1/2,0}- V^{1/2,0}\right)
+12 \left(V_{m_H}^{1/2,1/2}- V^{1/2,1/2}\right)\;,
\end{eqnarray}
which, together with (\ref{expansion}) and 
$m_H^2 = -\lambda\mu^2/(\Lambda^2V)$, yields
\begin{equation}
V_H = -\frac{1}{720\pi}\frac{\lambda\mu^2}{\Lambda^2 V}
\left( - 5\frac{R_2}{R_1^3} + \frac{5}{2}\frac{R_1}{R_2^3} + \ldots\right)\;.
\end{equation} 
For the {\bf 10}-plets $H_{3,4}$ the choice of parities is different, leading
to color triplets as zero modes. The corresponding Casimir energy is given by
\begin{eqnarray}
V'_H &=&
12 \left(V_{m_H}^{0,0}- V^{0,0}\right)
+ 12 \left(V_{m_H}^{0,1/2}- V^{0,1/2}\right) \nonumber \\
&&+ 8 \left(V_{m_H}^{1/2,0}- V^{1/2,0}\right)
+8  \left(V_{m_H}^{1/2,1/2}- V^{1/2,1/2}\right) \;.
\end{eqnarray}
Here we have neglected the supersymmetric brane masses (cf.~\cite{abc03})
which cancel in the behavior at small $R_{1,2}$,
\begin{equation}
V'_H = -\frac{1}{720\pi}\frac{\lambda'\mu^2}{\Lambda^2 V}
\left(10\frac{R_2}{R_1^3} + \frac{5}{2}\frac{R_1}{R_2^3} + \ldots\right)\;.
\end{equation} 
In the same way one obtains for the {\bf 16}-plets,
\begin{eqnarray}
V_{\Phi} &=&
2 \left(V_{M'}^{0,0}- V^{0,0}_M\right)
+ 16 \left(V_{M'}^{0,1/2}- V^{0,1/2}_M \right)
+ 24  \left(V_{M'}^{1/2,1/2}- V^{1/2,1/2}_M\right)
 \nonumber \\
&&+ 8 \left(V_{m_{\Phi}}^{1/2,0}- V^{1/2,0}\right)
+14 \left(V_{m_{\Phi}}^{0,0}- V^{0,0}\right)\;,
\end{eqnarray}
with $M'=\sqrt{M^2+m_{\Phi}^2}$, which yields for small radii
\begin{equation}
V_{\Phi} = -\frac{1}{720\pi}\frac{\lambda''\mu^2}{\Lambda^2 V}
\left(4\frac{R_2}{R_1^3} - 11\frac{R_1}{R_2^3} + \ldots\right)\;.
\end{equation}   

The four contributions to the Casimir energy, $V_g$, $V_H$, $V_H'$ and
$V_\Phi$ are displayed in Figure~2. 
Note that features at larger radii, like the profile
in the $R_2$-direction for $V_{\Phi}$, can be lost
in the simplified expression where we keep only the leading term in $\mu^2$.
The behavior at small radii however is unchanged and obvious
from the analytic expressions given above. Note that only $V_H'$ is repulsive
in all directions at small radii.

To leading order in $1/\Lambda$, the Casimir energy is determined by the
contribution from hypermultiplets since the gaugino mass is stronger volume
suppressed than the scalar masses. Depending on signs and magnitude of 
$\lambda$, $\lambda'$ and $\lambda''$, the resulting behavior at small radii 
can be attractive or repulsive. As an example, we shall assume in the following
$\lambda' <0, |\lambda'| \gg |\lambda|, |\lambda''|$ which yields a repulsive behavior at
small radii.

\section{Stabilization of the Compact Dimensions}

In the previous section we have calculated quantum corrections to the effective
potential at small radii and we have seen that, depending on the supersymmetry 
breaking parameters, the behavior can be attractive or repulsive. In the 
latter case a bulk cosmological term can lead to stabilization of the compact
dimensions \cite{pp01}. As we shall show in this section, stabilization can 
also follow from the interplay of the Higgs mechanism in 6D, which generates 
bulk mass terms, and supersymmetry breaking on the brane.

Consider the mass $M$ generated by spontaneous symmetry breaking
as discussed in Section~2 (cf.~(\ref{higgsmass})),
\begin{equation}\label{mass}
M^2 \simeq g_6^2 \langle\Phi^c\rangle^2 = g_4^2 V\langle\Phi^c \rangle^2\;,
\end{equation}
where $g_6$ has dimension length and $g_4 = g_6/\sqrt{V}$ is dimensionless.
For simplicity, we shall assume that $M$ is small compared to the Kaluza-Klein
masses and approximately constant.

In orbifold compactifications of the heterotic string expectation values
$\langle \Phi \rangle$ can be induced by localized Fayet-Iliopoulos 
terms. Vanishing of the D-terms then implies
\begin{equation}
V \langle \Phi^c \rangle^2 = C \Lambda^2\;,
\end{equation}
where $C \ll 1$ is a loop factor and $\Lambda$ is the string scale or, more 
generally, the UV cutoff of the model. For instance, in the 6D model of 
\cite{bls07} one finds for the localized anomalous $U(1)$'s,
$C\Lambda^2 \sim g M_{\mathrm{P}}^2/(384\pi^2)$.

Supersymmetry breaking by a brane field $S$, with $\mu = F_S/\Lambda$\;,
leads to a `classical' vacuum energy density,
\begin{eqnarray}
V^{(0)} &=& - \lambda''\int d^2y \int d^4\theta\frac{1}{\Lambda^4}\delta^2(y) 
\langle S^\dagger S(\Phi^\dagger\Phi +\Phi^{c\dagger}\Phi^c)\rangle \simeq 
- \lambda''\ \frac{\mu^2}{\Lambda^2}\ \langle\Phi^c\rangle^2 \nonumber\\
&=& - \lambda''\ \frac{\mu^2 C}{V}\;,
\end{eqnarray} 
with $V = (2\pi)^2 R_1R_2$.
For $\lambda''>0$, $V^{(0)}$ is attractive at large radii. Note that this
supersymmetry breaking mass term does not lead to a negative mass squared for 
$\Phi$ and $\Phi^c$ since these fields are assumed to be stabilized by much 
larger supersymmetry preserving masses at the minimum.
We assume that no tachyonic mass terms are generated for fields
whose expectation values are not fixed by the D--term potential.

The classical energy density $V^{(0)}$ together with 
the Casimir energy $V^{(1)} = V_H'$ yields the total energy density,
\begin{eqnarray}
V_\mathrm{tot}(R_1,R_2) &=& V^{(0)}(R_1,R_2) + V^{(1)}(R_1,R_2) \nonumber\\ 
&=& - \frac{1}{288\pi^3}\frac{\mu^2\lambda'}{\Lambda^2}\left(\frac{1}{R_1^4}
+ \frac{1}{4 R_2^4}\right) - \frac{\lambda''}{4\pi^2}\frac{\mu^2C}{R_1R_2}\;.
\end{eqnarray} 
The effective potential is attractive at large radii and, for $\lambda' < 0$,
i.e. $m^2_{H_{3,4}} > 0$, repulsive at small radii.
One easily verifies that the effective potential 
$V_\mathrm{tot}$ has a stable minimum at
\begin{equation}
R_1^\mathrm{min} = \sqrt{2} R_2^\mathrm{min}\;, 
\quad R_2^\mathrm{min} = \frac{2^{1/4}}{12\sqrt{\pi}}
\sqrt{\frac{-\lambda'}{\lambda''}}\frac{1}{M}\;.
\end{equation}
Here $M$ is the mass given by Eq.~(\ref{mass}) at the minimum, and we have
assumed $g_4(V_\mathrm{min})\simeq 1/\sqrt{2}$, as it is the case for standard
model gauge interactions. As Figure~3 illustrates, the total energy density 
$V_\mathrm{tot}$ is very flat for large radii. 

\begin{figure}
  \centering
    \includegraphics[width=12cm,height=8cm]{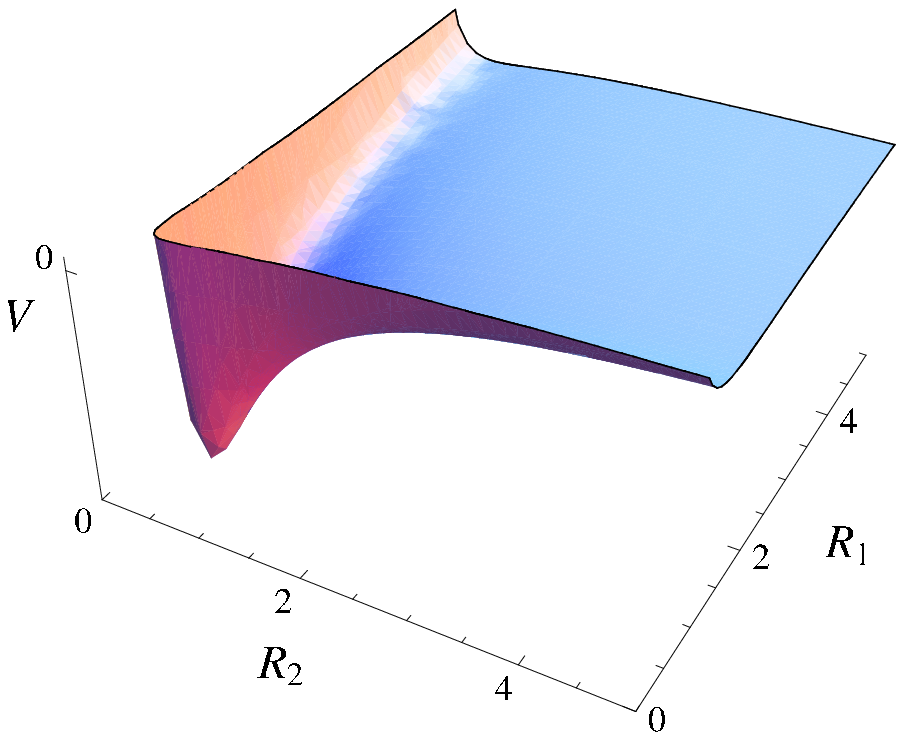}  
  \caption{Casimir energy of the $\bf 10$-plets $H_3$ and $H_4$ together with
  the classical energy density from the supersymmetry breaking brane.}
  \label{fig:potential}
\end{figure}

In orbifold compactifications of the heterotic string one typically has 
$M \sim M_\mathrm{GUT}$. It is very remarkable that the interplay
of gauge and supersymmetry breaking has lead to a stabilization at
$R^\mathrm{min} \sim 1/M_\mathrm{GUT}$, independent of the scale $\mu$ 
of supersymmetry breaking. The reason is that both, the classical vacuum 
energy density as well as the one-loop Casimir energy are proportional to 
$\mu^2$ which therefore does not affect the position of the minimum. 
Another interesting implication of the potential is that for $\mu \ll M_\mathrm{GUT}$,
\begin{equation}
\Delta V_\mathrm{tot}(R^\mathrm{min}) = V_\mathrm{tot}(\infty) - V_\mathrm{tot}(R^\mathrm{min}) \ 
\sim\ \mu^2  M_\mathrm{GUT}^2\ \ll\ M_\mathrm{GUT}^4\;.
\end{equation} 
Note that the energy density $V_{\mathrm{tot}}$ is negative at the minimum. It has to be
tuned to zero by means of a brane cosmological constant.
In a full supergravity treatment of stabilization also the interactions
of the supersymmetry breaking brane field with the radion fields have
to be taken into account.

The fact that the energy density difference $V_\mathrm{tot}(\infty)-V_\mathrm{tot}(R^\mathrm{min})$ 
is much smaller than $M_\mathrm{GUT}^4$ has important
cosmological consequences. In the thermal phase of the early universe, the
volume of the compact dimensions and, correspondingly, the value of 4D
coupling constants begins to change already at temperatures 
$T \sim \sqrt{\mu M_\mathrm{GUT}} \ll M_\mathrm{GUT}$ (cf.~\cite{bhx042}).

\section{Conclusions}

We have calculated the one-loop Casimir energy for bulk fields on the orbifold
$T^2/\mathbbm{Z}_2^3$. As expected, depending on the boundary conditions, 
the behavior at small radii can be attractive or repulsive. For the considered
supersymmetric model, the Casimir energy is proportional to the scale of
supersymmetry breaking. The relative strength of the couplings of the different
bulk fields to the supersymmetry breaking brane field then determines whether 
the behavior of the total energy density is repulsive or attractive at
small radii. 

Quantum corrections also modify the behavior at large radii. In orbifold
compactifications with $U(1)$ gauge factors, generically Fayet-Iliopoulos
terms are generated locally at the orbifold fixed points. This leads to
a breaking of these $U(1)$ gauge symmetries by the Higgs mechanism. Since
the symmetry breaking is induced by local terms, the generated masses
scale like $M \sim 1/\sqrt{V}$ with the volume of the compact dimensions.

The coupling of the bulk Higgs field to the supersymmetry breaking brane
field gives rise to a classical contribution to the total energy density
which scales like $1/V$ with the volume. Depending on the sign of the coupling,
the behavior of the energy density at large radii can be attractive or
repulsive. An attractive behavior at large radii, together with a repulsive
behavior due to the Casimir energy at small radii, can stabilize the 
compact dimensions. Since the supersymmetry breaking scale factorizes, the 
vacuum size of the compact dimensions is determined by the remaining mass 
scale, the mass $M$ generated by the Higgs mechanism, 
$R^\mathrm{min} \sim 1/M \sim 1/M_\mathrm{GUT}$.
At the minimum the energy density  $V_{\mathrm{tot}}$ is negative and has to be tuned to zero
by adding a brane cosmological term.

The characteristic feature of the described stabilization mechanism is a 
potential well much smaller than the GUT scale, 
$\Delta V_\mathrm{tot}(R^\mathrm{min}) \sim\ \mu^2  M_\mathrm{GUT}^2\ \ll\ M_\mathrm{GUT}^4$.
Clearly, this has important
cosmological consequences, both for the thermal phase of the early universe
as well as a possible earlier inflationary phase.

\section*{Acknowledgments}
We would like to thank
L.~Covi, K.~Fredenhagen, G.~von~Gersdorff, A.~Hebecker, J.~M\"oller, 
S.~Parameswaran, M.~Peloso, E.~Poppitz, M.~Ratz and J.~Schmidt for 
helpful discussions.
This work has been supported by the SFB-Transregio 27 ``Neutrinos and 
Beyond'' and by the DFG cluster of excellence ``Origin and 
Structure of the Universe''.

\newpage

\appendix

\section{Mode Expansion on $T^2/{\mathbbm Z}_2^3$} 

The orbifold $T^2/\mathbbm{Z}_2^3$ has four fixed points which we denote by 
$y_\mathrm{O}=(0,0)$, $y_\mathrm{PS}=(\pi R_1/2,0)$, 
$y_\mathrm{GG} = (0,\pi R_2/2)$ and $y_\mathrm{fl} = (\pi R_1/2,\pi R_2/2)$ 
(cf.~\cite{abc022}). The possible boundary conditions of functions on this
orbifold are characterized by three parities,
($a,b = +,-$),
\begin{eqnarray}\label{boundary} 
\phi_{\pm a b}(y_\mathrm{O}-y) &=& \pm \phi_{\pm a b}(y_\mathrm{O}+y)\;,
\nonumber\\
\phi_{a \pm b}(y_\mathrm{PS}-y) &=& \pm \phi_{a \pm b}(y_\mathrm{PS}+y)\;,
\nonumber\\
\phi_{a b \pm}(y_\mathrm{GG}-y) &=& \pm \phi_{a b \pm}(y_\mathrm{GG}+y)\;.
\end{eqnarray}
It is straightforward to define an orthonormal basis on the torus.
The mode expansion of functions with the boundary conditions (\ref{boundary})
then reads explicitly,
\begin{subequations}
\begin{align}
\phi_{+++}(x,y) &=\frac{1}{\sqrt{2 \pi^2 R_1R_2 2^{\delta_{n,0}\delta_{m,0}}}}
 \left[ \delta_{0,m} \sum_{n=0}^{\infty}+
\sum_{m=1}^{\infty}\sum_{n=-\infty}^{\infty} \right]
\phi_{+++}^{(2m,2n)}(x)\nonumber \\
&\hspace{6cm} \times \cos\left(\frac{2my_1}{R_1}+\frac{2ny_2}{R_2}\right) , 
 \\
\phi_{++-}(x,y) &=\frac{1}{\sqrt{2 \pi^2 R_1R_2}}
 \left[ \delta_{0,m} \sum_{n=0}^{\infty}+
\sum_{m=1}^{\infty}\sum_{n=-\infty}^{\infty} \right]
\phi_{++-}^{(2m,2n+1)}(x)\nonumber \\
&\hspace{4.95cm} \times\cos\left(\frac{2my_1}{R_1}+\frac{(2n+1)y_2}{R_2}\right) ,
\\ 
\phi_{+-+}(x,y) &=\frac{1}{\sqrt{2 \pi^2 R_1R_2}}
 \left[ 
\sum_{m=0}^{\infty}\sum_{n=-\infty}^{\infty} \right]
\phi_{+-+}^{(2m+1,2n)}(x)\nonumber \\
&\hspace{4.63cm} \times\cos\left(\frac{(2m+1)y_1}{R_1}
+\frac{(2n)y_2}{R_2}\right) ,
\\
\phi_{+--}(x,y) &=\frac{1}{\sqrt{2 \pi^2 R_1R_2}}
 \left[
\sum_{m=0}^{\infty}\sum_{n=-\infty}^{\infty} \right]
\phi_{+--}^{(2m+1,2n+1)}(x)\nonumber \\
&\hspace{3.93cm} \times\cos\left(\frac{(2m+1)y_1}{R_1}
+\frac{(2n+1)y_2}{R_2}\right) ,
\\
\phi_{-++}(x,y) &=\frac{1}{\sqrt{2 \pi^2 R_1R_2}}
 \left[
\sum_{m=0}^{\infty}\sum_{n=-\infty}^{\infty} \right]
\phi_{-++}^{(2m+1,2n+1)}(x)
\nonumber \\
&\hspace{3.96cm} \times
\sin\left(\frac{(2m+1)y_1}{R_1}+\frac{(2n+1)y_2}{R_2}\right) ,\\
\phi_{-+-}(x,y) &=\frac{1}{\sqrt{2 \pi^2 R_1R_2}}
 \left[ 
\sum_{m=0}^{\infty}\sum_{n=-\infty}^{\infty} \right]
\phi_{-+-}^{(2m+1,2n)}(x)
\nonumber \\
&\hspace{5cm} \times
\sin\left(\frac{(2m+1)y_1}{R_1}+\frac{2n y_2}{R_2}\right) ,\\
\phi_{--+}(x,y) &=\frac{1}{\sqrt{2 \pi^2 R_1R_2}}
 \left[ \delta_{0,m} \sum_{n=0}^{\infty}+
\sum_{m=1}^{\infty}\sum_{n=-\infty}^{\infty} \right]
\phi_{++-}^{(2m,2n+1)}(x)
\nonumber \\
&\hspace{5cm} \times
\sin\left(\frac{2my_1}{R_1}+\frac{(2n+1)y_2}{R_2}\right) ,\\
\phi_{---}(x,y) &=\frac{1}{\sqrt{2 \pi^2 R_1R_2}}
 \left[ \delta_{0,m} \sum_{n=0}^{\infty}+
\sum_{m=1}^{\infty}\sum_{n=-\infty}^{\infty} \right]
\phi_{---}^{(2m,2n)}(x)
\nonumber \\
&\hspace{5.7cm} \times
\sin\left(\frac{2my_1}{R_1}+\frac{(2n)y_2}{R_2}\right) .
\end{align}
\end{subequations}

\section{Evaluation of Casimir Sums}

Our evaluation of the Casimir double sums requires two single sums which
we shall now consider. 
The first sum reads
\begin{align}
 \widetilde{F}(s;a,c) 
\equiv \sum_{m=0}^{\infty}\frac{1}{\left[(m+a)^2+c^2\right]^{s}}\;.
\end{align}
This is a series of the generalized Epstein-Hurwitz zeta type. 
The result can be found in \cite{eli94} 
and is given by
\begin{align}
 \widetilde{F}(s;a,c) =& \frac{c^{-2s}}{\Gamma(s)}
\sum_{m=0}^{\infty}\frac{(-1)^m\Gamma(m+s)}{m!}c^{-2m}\zeta_H(-2m,a)
+ \sqrt{\pi}\frac{\Gamma(s-\tfrac{1}{2})}{2\Gamma(s)}c^{1-2s}\nonumber \\
&+\frac{2\pi^s}{\Gamma(s)}c^{1/2-s}\sum_{p=1}^{\infty}p^{s-1/2}\cos(2\pi p a)
K_{s-1/2}(2\pi p c) \;,
\end{align}
where $\zeta_H(s,a)$ is the Hurwitz zeta-function.
Note that this is not a convergent series but an asymptotic one.
In the following it will be important that
$\zeta_H(-2m,0)=\zeta_H(-2n,1/2)=0$ for $m\in \mathbbm{N}$ and $n\in 
\mathbbm{N}_0$. In our case, the first sum in $\widetilde{F}(s;a,c)$ thus 
reduces to a single term. For $a=1/2$ the sum vanishes, and for $a=0$ only
the first term contributes; with $\zeta_H(0,0)=1/2$ one obtains $c^{-2s}/2$.

The second, related sum is given by
\begin{align}
\label{SingleSum}
F(s;a,c) \equiv 
\sum_{m=-\infty}^\infty\frac{1}{\left[(m + a)^{2} + c^{2}\right]^s}\;. 
\end{align}
Using the two identities ($m\in \mathbbm{N}$)
\begin{align}
\zeta_H(-2m,a) &=-\zeta_H(-2m,1-a) \;, \\ 
F(s;a,c) &= \widetilde{F}(s;a,c) + \widetilde{F}(s;1-a,c)\;,
\end{align}
one easily obtains, in agreement with \cite{pp01},
\begin{equation}
F(s;a,c) = \frac{\sqrt{\pi}}{\Gamma(s)} |c|^{1-2s} 
\left[\Gamma\left(s-\tfrac{1}{2}\right)
+ 4\sum_{p=1}^{\infty} \cos(2\pi p a)(\pi\,p\,|c|)^{s -\frac{1}{2}} 
K_{s -\frac{1}{2}} (2\pi \,p \,|c|) \right] \,.
\end{equation}
These two sums provide the basis for our evaluation of the Casimir sums.

\subsection[Casimir Sum (I) on $T^2/{\mathbbm Z}_2^3$] 
{Casimir Sum (I) on ${ \bf T^2/\mathbbm Z}_2^3$} 

We first consider the summation
\begin{align}
 \left[\sum \right]_{m,n} = 
\sum_{m=0}^{\infty}\sum_{n=-\infty}^{\infty} \;.  
\end{align}
In this case the Casimir energy (cf.~(\ref{zeta})) is obtained from 
\begin{align}
\sum_{m=0}^{\infty}\sum_{n=-\infty}^{\infty}
\left[e^2(m+\alpha)^2 +(n+\beta)^2 + \kappa^2\right]^{-s} \;, 
\end{align}
where we have shifted $s\to s+2$ and defined $\kappa^2= \tfrac{R_2^2}{4}M^2$.
Using the expression for $F(s;a,c)$, we can perform the sum over $n$,
\begin{align}
&\sum_{m=0}^{\infty}\sum_{n=-\infty}^{\infty}
\left[e^2(m+\alpha)^2
+(n+\beta)^2 + \kappa^2\right]^{-s}  \nonumber \\
&=
\sqrt{\pi} \frac{ \Gamma(s-\frac{1}{2}) }{\Gamma(s)} 
 \sum_{m=0}^{\infty}(e^2(m+\alpha)^2 +\kappa^2)^{1/2-s} 
 \nonumber \\
&\quad +\frac{4\sqrt{\pi}}{\Gamma(s)} \sum_{p=1}^{+\infty} \cos(2 \pi p \beta) 
 \sum_{m=0}^{\infty}(\pi \,p)^{s-\tfrac{1}{2}}
\left(\sqrt{e^2(m+\alpha)^2 +\kappa^2}\right)^{\frac{1}{2}-s}  \nonumber \\
& \hspace{6cm}
K_{s -\frac{1}{2}} (2\pi \,p \,\sqrt{e^2(m+\alpha)^2 +\kappa^2}) \nonumber \\
&\equiv \  f_1(s) + f_2(s) \;. 
\end{align}
Let us consider $f_1(s)$ first. The sum over $m$ can be performed with the
help of $\widetilde{F}(s;a,c)$,
\begin{align}
f_1(s)
=&
\sqrt{\pi} \frac{ \Gamma(s-\frac{1}{2}) }{\Gamma(s)} 
\sum_{m=0}^{\infty}(e^2(m+\alpha)^2 +\kappa^2)^{1/2-s} 
 \nonumber \\
=& \sqrt{\pi}\frac{\Gamma(s-1/2)}{\Gamma(s)}\kappa^{1-2s}
\zeta_H(0,\alpha) + \frac{\pi}{2(s-1)}
\frac{\kappa^{2-2s}}{e}\nonumber \\
&+\frac{2\pi^{s}}{\Gamma(s)} e^{-s}\kappa^{1-s}
\sum_{p=1}^{\infty}p^{s-1}\cos(2\pi p \alpha)
K_{s-1}(2\pi p \left(\tfrac{\kappa}{e}\right)) \;.
\end{align}
Recalling the shift in $s$,
we can now write $\zeta(s)$ (\ref{zeta}) as 
\begin{align}
\zeta(s)=& \frac{1}{32\pi^2}\left(\frac{4}{R_2^2}\right)^{-s} 
\frac{\mu_r^{2s+4}}{s(s+1)}
\bigg\{ \sqrt{\pi}\frac{\Gamma(s-1/2)}{\Gamma(s)} \kappa^{1-2s}
\zeta_H(0,\alpha) + \frac{\pi}{2(s-1)}
\frac{\kappa^{2-2s}}{e}\nonumber \\
&+\frac{2\pi^{s}}{\Gamma(s)} e^{-s}\kappa^{1-s}
\sum_{p=1}^{\infty}p^{s-1}\cos(2\pi p \alpha)
K_{s-1}(2\pi p \left(\tfrac{\kappa}{e}\right)) 
 \nonumber \\
&+  \frac{4\sqrt{\pi}}{\Gamma(s)} \sum_{p=1}^{+\infty} \cos(2 \pi p \beta) 
 \sum_{m=0}^{\infty}(\pi \,p)^{s-\tfrac{1}{2}}
\left(\sqrt{e^2(m+\alpha)^2 +\kappa^2}\right)^{\frac{1}{2}-s} \nonumber \\
&\hspace{5cm} 
K_{s -\frac{1}{2}} (2\pi \,p \,\sqrt{e^2(m+\alpha)^2 +\kappa^2})\bigg\}\;. 
\end{align}
Now we have to differentiate with respect to $s$ and set $s=-2$.
Since $\Gamma(-2)= \infty$, the derivative has only to act on $\Gamma(s)$
if the corresponding term is inversely proportional to $\Gamma(s)$.
Performing the differentiation, using
\begin{align}
\frac{\text{d}}{\text{d}s}\frac{1}{\Gamma(s)}\bigg|_{s=-2}
=-\frac{\Gamma^{'}(s)}{\Gamma(s)^2}\bigg|_{s=-2} = 2 \;, 
\end{align}
and $K_{a}(z)=K_{-a}(z)$, and substituting
again $e=R_2/R_1$ and $\kappa^2=R_2M/2$, we finally obtain for the
Casimir energy,
\begin{align}
V_{\text{M}}^{\alpha,\beta (I)}
=&\ \frac{M^5R_2}{120\pi} \zeta_H(0,\alpha)
+ \frac{M^6 R_1 R_2}{768\pi}
 \left(\frac{11}{12} - \log\left(\frac{M}{\mu_r}\right)\right) \nonumber \\
&-\frac{1}{8\pi^4} \frac{M^3R_2}{R_1^2}
\sum_{p=1}^{\infty}\frac{\cos(2\pi p \alpha)}{p^3}
K_{3}(\pi p M R_1) \nonumber \\
&- \frac{2}{\pi^4} \frac{1}{R_2^4}
\sum_{p=1}^{\infty} \frac{\cos(2 \pi p \beta)}{p^{5/2}} 
 \sum_{m=0}^{\infty}
\left(\frac{R_2}{R_1}\sqrt{(m+\alpha)^2 +\tfrac{M^2R_1^2}{4}}
\right)^{\frac{5}{2}} \nonumber \\
&\hspace{2cm}K_{5/2} \left(2\pi \,p \,\tfrac{R_2}{R_1}
\sqrt{(m+\alpha)^2 + M^2R_1^2/4}\right) \;.
\end{align}
The second term corresponds to a finite part of the 
6D cosmological constant. The dependence on the regularization scale $\mu_r$
shows that an infinite contribution has been subtracted.

\subsection[Casimir Sum (II) on $T^2/{\mathbbm Z}_2^3$] 
{Casimir Sum (II) on ${ \bf T^2/\mathbbm Z}_2^3$} 

The second relevant summation is
\begin{align}
 \left[\sum \right]_{m,n} = 
 \left[ \delta_{m,0} \sum_{n=0}^{\infty}+
\sum_{m=1}^{\infty}\sum_{n=-\infty}^{\infty} \right] \;.
\end{align}
For the corresponding boundary conditions one has $\alpha=0$.
The Casimir sum can then be written as 
\begin{align}
& \left[ \delta_{m,0} \sum_{n=0}^{\infty}+
\sum_{m=1}^{\infty}\sum_{n=-\infty}^{\infty} \right]
\left[e^2 m^2 + (n+\beta)^2 + \kappa^2\right]^{-s}   \nonumber \\
=\ & \left[\delta_{m,0}\sum_{n=0}^{\infty}
+\sum_{m=0}^{\infty}\sum_{n=-\infty}^{\infty} 
-\ \delta_{m,0}\sum_{n=-\infty}^{\infty}\right]
\left[e^2 m^2 +(n+\beta)^2 + \kappa^2\right]^{-s} \;,
\end{align}
where we again shifted $s\to s+2$ and set $\tfrac{R_2^2}{4}M^2 = \kappa^2$.
The double sum is the sum (I) which we have already evaluated. Using
\begin{align}
\sum_{n=-\infty}^{-1}\left[(n+\beta)^2+\kappa^2\right]^{-s} = 
\sum_{n=0}^{\infty}\left[(n+1-\beta)^2+\kappa^2\right]^{-s}\;.
\end{align}
one easily finds for the remaining piece\footnote{
Note that $\zeta_H(0,1)=-1/2$, and $\zeta_H(-2m,1)=0$ for $m\in \mathbbm{N}$.}
\begin{align}
f_3(s)&=- \sum_{n=0}^{\infty}\left[(n+1-\beta)^2
+\kappa^2\right]^{-s}\nonumber \\
&= -\kappa^{-2s}\zeta_H(0,1-\beta)
- \sqrt{\pi}\frac{\Gamma(s-\tfrac{1}{2})}{2\Gamma(s)}
\kappa^{1-2s}\nonumber \\
&-\frac{2\pi^s}{\Gamma(s)}\kappa^{1/2-s}
\sum_{p=1}^{\infty}p^{s-1/2}\cos(2\pi p (1-\beta))
K_{s-1/2}(2\pi p \kappa) \;.
\end{align}
Differentiating the corresponding contribution to $\zeta(s)$, 
setting $s=-2$,
and substituting $\kappa=R_2M/2$ yields the Casimir energy,
\begin{eqnarray}
V_M^{0,\beta (II)}
&=& V_M^{0,\beta (I)} \nonumber \\
&& +\frac{M^4}{64\pi^2} 
\left(\frac{3}{2} - 2\log\left(\frac{M}{\mu_r}\right)\right)\zeta_H(0,1-\beta)
- \frac{1}{240\pi} M^5 R_2 \nonumber \\
&& - \frac{1}{\pi^4}\frac{1}{R_2^4}
\sum_{p=1}^{\infty}\frac{\cos(2\pi p (1-\beta))}{p^{5/2}}
\left(\frac{MR_2}{2}\right)^{5/2}K_{5/2}\left(\pi p MR_2 \right)\;.
\end{eqnarray}
The first of the additional terms does not depend on the radii. It represents
a finite contribution to the brane cosmological term. The dependence on
the regularization scale $\mu_r$ again shows that a divergent contribution
has been subtracted.

\end{document}